\documentclass[%
superscriptaddress,
twocolumn,
groupedaddress,
showpacs,
amsmath,amssymb,
aps,
prb,
floatfix,
showkeys,
reprint,
]{revtex4-1}

\usepackage{graphicx}% Include figure files
\usepackage{dcolumn}% Align table columns on decimal point
\usepackage{bm}% bold math
\usepackage{amsmath}
\usepackage{amssymb}
\usepackage{multirow}
\usepackage{color,colortbl}
\usepackage{booktabs}
\usepackage{tabularx}
\usepackage[english]{babel}
\usepackage[utf8]{inputenc}
\usepackage{enumerate}
\usepackage{newunicodechar}
\newunicodechar{ﬁ}{fi}
\newunicodechar{ﬀ}{ff}
\usepackage{float}

\newcommand{\etal}{\textit{et al.\/}}
\newcommand{\ie}{i.\,e.,}
\newcommand{\eg}{e.\,g.,}
\newcommand{\uB}{\ensuremath{\mu_{\text{B}}}}
\newcommand{\eF}{\ensuremath{E_{\text{F}}}}

\def\sjh[#1]{\textcolor{blue}{#1}}
\begin{document}
%Reduce the space before and after an equation
%\abovedisplayskip=10pt
%\belowdisplayskip=10pt
%\abovedisplayshortskip=0pt
%\belowdisplayshortskip=7pt

%\title{Spin-orbit driven tunneling anisotropic magnetoresistance effect of Pb and Bi adatoms and dimers on Mn/W(110) substrate}
\title{Tunneling anisotropic magnetoresistance of Pb and Bi adatoms and dimers on Mn/W(110): A first-principles study}

\author{Soumyajyoti Haldar}
\email[Corresponding author: ]{haldar@physik.uni-kiel.de\\haldar.physics@gmail.com}
\affiliation{Institute of Theoretical Physics and Astrophysics, University of Kiel, Leibnizstrasse 15, 24098 Kiel, Germany}

\author{Mara Gutzeit}
\affiliation{Institute of Theoretical Physics and Astrophysics, University of Kiel, Leibnizstrasse 15, 24098 Kiel, Germany}

\author{Stefan Heinze}
\affiliation{Institute of Theoretical Physics and Astrophysics, University of Kiel, Leibnizstrasse 15, 24098 Kiel, Germany}

\date{\today}

\begin{abstract}
We show that Pb and Bi adatoms and dimers have a large tunneling anisotropic magnetoresistance (TAMR) of up to 60\% when adsorbed on a 
magnetic 
%SH: I optimized the abstract 
transition-metal surface due to strong spin-orbit coupling and the hybridization of $6p$ orbitals with $3d$ states of the magnetic layer. 
Using density functional theory, we have explored the TAMR effect of Pb and Bi adatoms and dimers adsorbed on a Mn monolayer on W(110). This surface
exhibits a noncollinear cycloidal spin spiral ground state with an angle of 173$^\circ$ between neighboring spins which allows to rotate the spin
quantization axis of an adatom or dimer quasi-continuously and is ideally suited to explore the angular dependence of TAMR using scanning tunneling 
microscopy (STM). We find that the induced magnetic moments of Pb and Bi adatoms and dimers are small, however, the spin-polarization of the 
local density of states (LDOS) is still very large. The TAMR obtained from the anisotropy of the vacuum LDOS is up to 50-60 \% for adatoms. 
For dimers the TAMR depends sensitively on the dimer orientation with respect to the crystallographic directions of the surface due to the
formation of bonds between the adatoms with the Mn surface atoms and the symmetry of the spin-orbit coupling induced mixing. 
Dimers oriented along the spin spiral direction of the Mn monolayer display the largest TAMR of 60 \% which is due to hybrid $6p-3d$ states
of the dimers and the Mn layer. 
%Investigating the anisotropy of the local density of states (LDOS) in the vacuum, we 
%show small induced moments, however, large spin polarization and large TAMR for both adatoms and for suitably oriented dimers,
%which can be measured in STM experiment. 
%Our calculation indicates that the hybridization with underlying Mn layers and $6p$-$3d$ hybrid states depend surprisingly on dimer orientation. 
\end{abstract}

\maketitle
\section{Introduction}
The tunneling magnetoresistance 
%SH
(TMR), in which the flow of current depends on the relative magnetization directions of two magnetic layers, 
has a significant 
%SH
impact on modern day applications ranging from spintronics to magnetic data storage.
Using spin-polarized scanning tunneling microscopy (STM), it is even  possible to detect the TMR effect for single magnetic adatoms on 
surfaces~\cite{Yayon2007,Meier82,Tao2009,Loth2010,Ziegler2011,Lazo2012,Khajetoorians55}. The resistance can also  depend on the magnetization direction relative to the current direction because of spin-orbit coupling (SOC), which is known as the tunneling anisotropic magnetoresistance (TAMR)~\cite{Bode2002,Gould2004}. 
The TAMR is driven by SOC which couples spin and orbital momentum degrees of freedom by the 
Hamiltonian $H_{SOC} = \xi \, \mathbf{L} \cdot \mathbf{S}$,  where $\xi$, $\mathbf{L}$, and $\mathbf{S}$ are the SOC constant, orbital momentum operator and spin operator, respectively. 
SOC and magnetocrystalline anisotropy effects depend on the environment of an adatom and hence can be tuned by adatom adsorption which have been studied quite extensively~~\cite{Gambardella1130,Hirjibehedin1199,Loth1628,Khajetoorians2011,Rau988}.
The TAMR can be observed with only one ferromagnetic electrode and it does not require any coherent spin-dependent transport. Hence, the TAMR is very attractive for spintronics 
applications~\cite{Fert2008,Sinova2012}. The TAMR was first observed for a double layer of Fe on W(110)~\cite{Bode2002}. Subsequently, the TAMR has been observed in various systems, {\eg} planar ferromagnetic surfaces~\cite{Shick2006,Chantis2007}, tunnel junctions~\cite{Gould2004,Matos2009a,Matos2009b,Gao2007}, mechanically controlled break junctions~\cite{Viret2006,Bolotin2006}. The observed values of TAMR in the above cases are $\approx$ 10\%. 
%SH add Wulfhekel PRB from last year, about 60% TAMR ?
Attempts have been made to increase the value of TAMR by using $3d$ or $5d$ elements, {\eg} using isolated adatoms
~\cite{Neel2013,Schoneberg2016}, bimetallic alloys~\cite{Shick2010} and with antiferromagnetic electrodes~\cite{Park2011}.
%SJH 
Recently, Herv\'{e} {\etal} have reported a TAMR of up to 30\% for Co films on Ru(0001) mediated by surface states~\cite{Herve2018b}.

Another approach to tune SOC is to use single atoms and dimers of $6p$ elements. The strength of SOC scales with atomic number ($Z$), principal quantum number ($n$), and orbital quantum number ($l$) as $\xi \propto Z^{4}n^{-3}l^{-2}$. Hence, $6p$ elements such as Pb and Bi have a higher SOC strength as compared to that of the $3d$ or $5d$ elements studied before. Further tuning of SOC can be achieved by reducing the high rotational symmetry of single atom, {\ie} by using dimers of these elements. The effect of strong SOC on unsupported $6p$ dimers has been discussed recently~\cite{Borisova2016a}. In 
an experimental and theoretical study, Sch\"{o}neberg {\etal}~\cite{Schoneberg2018} have achieved TAMR values of $\approx$ 20\% by using suitably oriented Pb dimers on the Fe bilayer  on W(110) substrate where magnetic domains with out-of-plane magnetization and domain walls with in-plane magnetization can be observed~\cite{Bode2002}.

In recent years noncollinear magnetic structures at transition-metal interfaces have gained popularity as promising candidates for spintronic applications due to their interesting dynamical and transport 
properties~\cite{Fert2013,Nagaosa2013}. A monolayer Mn grown on W(110) surface (Mn/W(110)) is a prominent example which exhibits a noncollinear magnetic structure with a cycloidal 173$^{\circ}$ spin-spiral ground state along the $[1\overline{1}0]$ direction~\cite{Bode2007} 
%SH
that is driven by the Dzyaloshinskii-Moriya interaction. 
Using this magnetic surface with a noncollinear spin structure, it is possible to control the spin direction of adsorbed Co adatoms due to local exchange coupling
%SH
which has been demonstrated in recent experiments using scanning tunneling microscopy (STM) by Serrate {\etal}~\cite{Serrate2010,Serrate2016}. 
%SH
The noncollinear spin state of the Mn monolayer is reflected due to hybridization even in the orbitals of the adsorbed Co adatom~\cite{Haldar2018}. 
The possibility of controlling the magnetization direction of an adatom on this surface without the presence of external magnetic field makes 
this system very promising for TAMR studies. Compared to the domain walls of Fe/W(110) used in previous studies~\cite{Schoneberg2018} the spin structure of 
this surface is known on the atomic scale and allows a quasi continuous rotation of the local spin quantization axis.
Recently, Caffrey {\etal} have predicted TAMR values up to 50\% for Ir adatoms, i.e.~a $5d$ transition metal, on Mn/W(110)~\cite{Caffrey2014a}, however, experimental evidence is missing. 

Here we have explored Pb and Bi adatoms and dimers on Mn/W(110) in order to explore the magnitude of TAMR and its dependence on the $6p$  element and atomic arrangement on the surface.
We have used first-principles density functional theory (DFT) calculations to investigate the adsorption of Pb and Bi adatoms and dimers on Mn/W(110) and studied their electronic and magnetic properties. 
The spin structure of Mn/W(110) is locally well approximated as a two-dimensional antiferromagnet \cite{Heinze2000}. 
We considered two limiting cases of spin directions which are possible due to the cycloidal nature and propagation direction of the spin spiral in the Mn layer: (i) a magnetization direction 
perpendicular to the surface (out-of-plane) and (ii) a magnetization direction pointing along the $[1\overline{1}0]$ direction (in-plane). Our results indicate that the adsorption of these adatoms facilitates local enhancement of SOC above the surface leading to very large values of the TAMR of 50\% to 60\% for adatoms. The orientation of Pb and Bi dimers is shown to be crucial in order
to achieve even larger TAMR values. This can be understood based on the symmetry of the matrix elements of the SOC Hamiltonian as well as the hybridization of $6p$ adsorbate with $3d$ substrate
states.

This paper is organized as follows. First, we briefly discuss the computational methods used in our calculations. Then we proceed to discuss the structural, electronic, and magnetic properties, 
as well as the TAMR of adatoms and the same for dimers in different orientations.
The TAMR effects are discussed focusing on the local density of states at the adsorbate atoms and the Mn layer and the vacuum density of states and interpreted based on a simplified model. 
We summarize our main conclusions in the final section. 

\section{Computational details}
\label{sec:compdet}
In this work we used first-principles calculations using a plane wave based DFT code \textsc{vasp}~\cite{vasp1,vasp2} within the projector augmented wave method (PAW)~\cite{blo,blo1}. For the exchange-correlation, we have used the generalized gradient approximation (GGA) of Perdew-Burke-Ernzerhof
(PBE)~\cite{PBE,PBEerr}. For SOC, we followed the methods described by Hobbs {\etal}~\cite{Hobbs2000}. We used a 450 eV energy cutoff for the plane wave basis set convergence. 
Structural relaxations are performed using 
a $6\times6\times1$ k-point Monkhorst-Pack mesh~\cite{Monkhorst1976}. The vacuum local density of states (LDOS) was calculated by placing an empty sphere at a specific height of 5.3 {\AA} above the adatoms onto which the LDOS was projected. For the calculation of electronic properties, magnetic properties and LDOS, we have used $20\times20\times1$ k-point Monkhorst-Pack mesh. 

\subsection{Structural details}
\label{subsec:structure}
\begin{figure}[hbt]
	\centering
	\includegraphics[scale=0.85,clip]{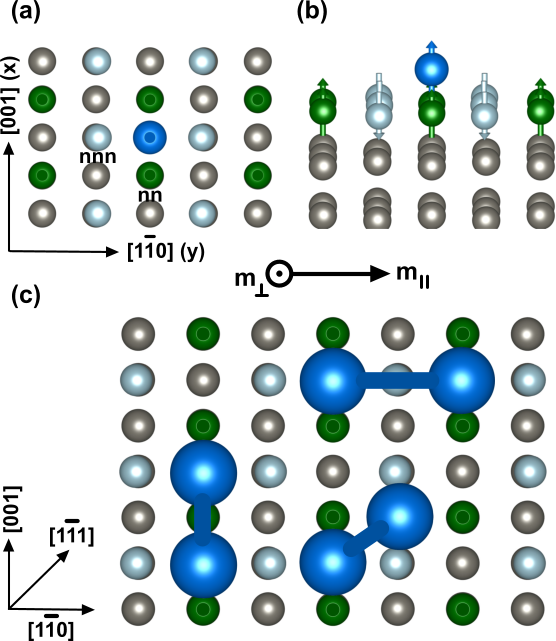}
	\caption{(a) Top view and (b) perspective view of a $c(4\times4)$ supercell used for the adatom on Mn/W(110) calculation. Gray spheres represent W atoms while Mn atoms are depicted as green (light blue) spheres with arrows showing the ferromagnetic (antiferromagnetic) magnetic moments with respect to the $6p$ adatom (dark blue sphere). `nn' and `nnn' are the nearest neighbor and next nearest neighbor Mn atoms to the adatom. $x$ and $y$ refer to the direction of coordinates for the supercell. (c) Top view of a $c(6\times6)$ supercell used for dimer adsorption on Mn/W(110) along with the three dimer orientations considered in our calculations. $m_\perp$ and $m_\parallel$ denote the direction of a perpendicular magnetization and a parallel magnetization with respect to the surface, respectively.}
	\label{fig:geom}
\end{figure}
We modeled Mn/W(110) using a symmetric slab consisting of five atomic layers of W with a pseudomorphic Mn layer on each side.
We have approximated the local magnetic order of the system as antiferromagnetic, {\ie} collinear due to the long periodicity of the spin spiral ground state \cite{Heinze2000,Bode2007,Serrate2010}.
The effect of the noncollinearity of the spin structure on the electronic states of adatoms has been studied before~\cite{Haldar2018}.
We used a $c(4\times4)$ AFM surface unit cell, as shown in Fig.~\ref{fig:geom}(a)-(b). The GGA calculated lattice constant of W, {\ie} 3.17 {\AA} is used for our calculations as it is in good agreement with the experimental value of 3.165 {\AA}. A thick vacuum layer of $\approx$ 25 {\AA} is included in the $z$ direction normal to the surface to remove interactions between repeating slabs. We added Pb or Bi adatoms at the hollow-site position on each Mn monolayer. The $c(4\times4)$ unit cell is large enough to keep the interactions between the periodic images of the adatoms small.
For the adsorption of dimers, we have used a larger $c(6\times6)$ AFM surface unit cell (see Fig.~\ref{fig:geom}(c)) to keep the interactions coming from the periodic images of the dimers negligible. In the case of Pb or Bi dimers, we have considered three possible dimer orientations on the surface: (i) along the [001] direction, (ii) along the $[1\overline{1}0]$ direction, and (iii) along the  $[1\overline{1}1]$ direction as shown in Fig.~\ref{fig:geom}(c).
%SH
The magnetization direction in calculations including SOC has been chosen normal to the surface, $\perp$, and along the $[1\bar{1}0]$ in-plane direction, $\parallel$, as enforced by the 
cycloidal nature of the underlying spin spiral structure of Mn/W(110) \cite{Bode2007}. 
The position of the adatoms, dimers and the Mn layers are relaxed with 0.01 eV/{\AA} force tolerance. We have kept the coordinates of W atoms fixed in all our calculations. 

\subsection{Tunneling anisotropic magnetoresistance}
\label{subsec:tamr}
Using the spectroscopic mode of an STM, the TAMR can be obtained by measuring the differential conductance (d$I$/d$V$) above an adatom or a dimer for two different magnetization directions. The TAMR is obtained from
\begin{align}
\mathrm{TAMR} &= \frac{[(\mathrm{d}I/\mathrm{d}V)_\perp-(\mathrm{d}I/\mathrm{d}V)_\parallel]}{(\mathrm{d}I/\mathrm{d}V)_\perp}\; ,
\end{align}
where $\perp$ and $\parallel$ denote a perpendicular magnetization and a parallel magnetization with respect to the surface, respectively. 
Within the Tersoff-Hamann model~\cite{Tersoff1983,Tersoff1985}, the d$I$/d$V$ signal is directly proportional to the local density of states (LDOS), 
$n(z, \epsilon)$, 
at the tip position in the vacuum, $z$, a few {\AA}ngstr\"{o}ms above the surface. Hence, the TAMR can be calculated theoretically from the anisotropy of 
the LDOS arising due to SOC~\cite{Bode2002,Neel2013}. Then the TAMR can be calculated as:
\begin{align}
\mathrm{TAMR} &= \frac{n_\perp(z, \epsilon)-n_\parallel(z, \epsilon)}{n_\perp(z, \epsilon)}\; .
\label{eq:TAMR}
\end{align}

\section{Results and Discussion}
\label{sec:results}

\subsection{Pb and Bi adatoms on Mn/W(110)}
\label{sbsec:adatoms}
\subsubsection{Structural and magnetic properties}
\label{sub2sec:str_mag}
\begin{table}[htb]
	\centering
	\caption{Relaxed distances (in {\AA}) of Pb and Bi 
	adatoms from the Mn atoms of the Mn/W(110) surface. $d_{\text{nn}}$ and $d_{\text{nnn}}$ denotes the nearest neighbor (nn) and the next-nearest neighbor (nnn) Mn atoms, respectively. $\Delta x$, $\Delta y$, and $\Delta z$ are the displacements with respect to the clean surface of Mn atoms after the adsorption of the adatoms. Positive (negative) values imply that the Mn atoms move towards (away from) the adatom.}
	\label{tab:distance}
    \begin{ruledtabular}
     \begin{tabular}{ c c c c c c c}
 & $d_{\text{nn}}$ & $d_{\text{nnn}}$ & $\Delta z_{\text{nn}}$ & $\Delta z_{\text{nnn}}$ & $\Delta x_{\text{nn}}$ & $\Delta y_{\text{nnn}}$ \\
\colrule
 Pb & 2.76 & 3.17 & $-0.13$ & +0.02 & $-0.01$ & +0.02 \\
 Bi & 2.70 & 2.97 & $-0.10$ & +0.06 & $-0.05$ & +0.03\\
\end{tabular}
\end{ruledtabular}
\end{table}
We begin our discussion with the local structural relaxations 
%SH
upon adsorption of Pb and Bi adatoms on Mn/W(110) which are tabulated in Table~\ref{tab:distance}. Our calculations indicate that the hollow site (see Fig.~\ref{fig:geom}(a)) is the most stable adsorption site for both Pb and Bi adatoms. The other sites, {\eg} bridge and top sites are unstable for both adatoms in our calculations and collapse to the hollow site position. 
The adsorption of the adatoms creates a buckling in the underlying Mn layer in the vicinity of the adsorption sites (see Table~\ref{tab:distance}). Significant changes can be observed for the nearest neighbor (nn) Mn atoms, which move away from the adatoms, while the next nearest neighbor (nnn) Mn adatoms move slightly towards the adatoms.

\begin{table}[htb]
	\centering
	\caption{Magnetic moments (in {\uB}) of the adsorbed Pb and Bi adatoms and the nearest neighbor (nn) and next nearest neighbor (nnn) Mn atoms of the Mn monolayer on W(110). For comparison the value of the clean Mn/W(110) surface is given.}
	\label{tab:magmom}
    \begin{ruledtabular}
	\begin{tabular}{c c c c c}
	 & Adatom & Mn$_{\text{nn}}$ & Mn$_{\text{nnn}}$ & Mn$_{\text{clean}}$ \\
\colrule
Pb & +0.00 & +2.36 & $-3.34$ & $\pm$3.41 \\
Bi & +0.08 & +2.60 & $-3.35$ & $\pm$3.41 \\
\end{tabular}
	\end{ruledtabular}
\end{table}
The magnetic properties of these systems are affected by the hybridization between the $6p$ adatoms and underlying Mn atoms of Mn monolayer (see Table~\ref{tab:magmom}). The clean Mn surface of Mn/W(110) have magnetic moments $\pm$ 3.41 {\uB}. The $p_z$ orbitals of the adatoms mainly hybridize with the $d_{z^2}$ orbitals of nn Mn adatoms. The magnetic moments of the nn Mn adatoms drop quite significantly for both atom types.  They are reduced by 1.05 {\uB} and 0.81 {\uB} for Pb and Bi adsorption, respectively. Due to the hybridization, the induced magnetic moment on Bi adatom is 0.08 {\uB}, whereas the Pb adatom is  non-magnetic.  

The effect of hybridization is less prominent for the nnn Mn adatoms where a slight reduction of magnetic moment $\sim$ 0.06 {\uB} occurs for both adatoms. 

\subsubsection{Electronic properties}
\begin{figure*}[htbp]
	\centering
	\includegraphics[scale=0.45,clip]{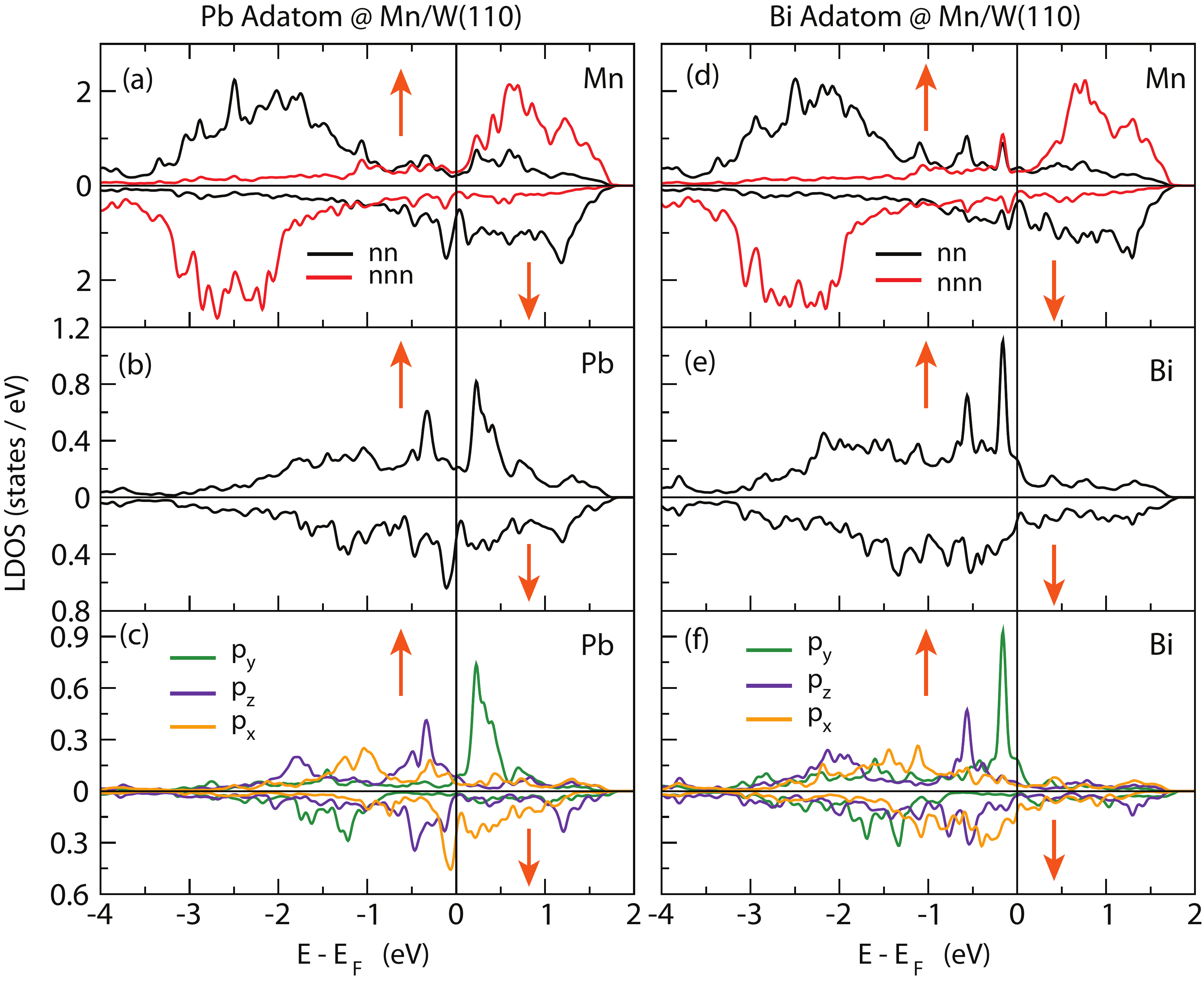}
	\caption{(a, d) Scalar relativistic spin-resolved LDOS of the neighboring (nearest `nn' and next-nearest `nnn') Mn atoms of the Pb and Bi adatom adsorbed on Mn/W(110). (b, e) Scalar relativistic spin-resolved LDOS of the Pb and Bi adatom. (c, f) Scalar relativistic $m_l$ decomposed spin-resolved LDOS for Pb and Bi adatom. Majority and minority states respectively are defined with respect to the nearest neighbor Mn atoms of the Mn monolayer. 
	The orange up and down arrow indicates majority and minority spin channels, respectively.}
	\label{fig:spinpol_adatom}
\end{figure*}
Next, we discuss the electronic properties of the $6p$ adatoms adsorbed on the Mn/W(110) surface. Fig.~\ref{fig:spinpol_adatom} shows the spin-resolved LDOS of the Pb and Bi adatom adsorbed on Mn/W(110), the LDOS of the neighboring Mn atoms, and $m_l$ decomposed $p$ states of Pb and Bi adatom. These calculations have been performed in the scalar relativistic approximation, {\ie} neglecting SOC. 

A possible hybridization can be observed by calculating and comparing the spin-resolved LDOS of the adatoms with the neighboring Mn states as shown in Fig.~\ref{fig:spinpol_adatom}. 
This hybridization effect is clearly observed %sticks out 
just below {\eF} where minority Mn peaks are located at the same position as $p_x$ and $p_z$ states of the adatoms.
Further interactions are observed 
%SH
for Pb
around {\eF}$-$0.50~eV, {\eF}$-$0.34~eV, where the states from the adatoms interact with the states from the nn Mn atoms. 
In this energy range one also sees reduced exchange splitting of the nn Mn states as compared to the nnn Mn states which
affects the magnetic moment of the nn Mn atoms as mentioned in section~\ref{sub2sec:str_mag}. 
The magnetic moment of nn Mn atoms drops to 2.36 {\uB} and  2.6 {\uB} upon adsorption of the Pb and Bi adatom, respectively.
Despite the small spin splitting observed for both adatoms, 
the spin polarization of the adatoms 
%SH
is quite large.
The spin polarization of the adatoms varies in-between $\pm 40$\% which is mainly arising from the $p_z$ and $p_y$ states of the adatoms. 

\begin{figure}[htbp]
	\centering
	\includegraphics[scale=0.85,clip]{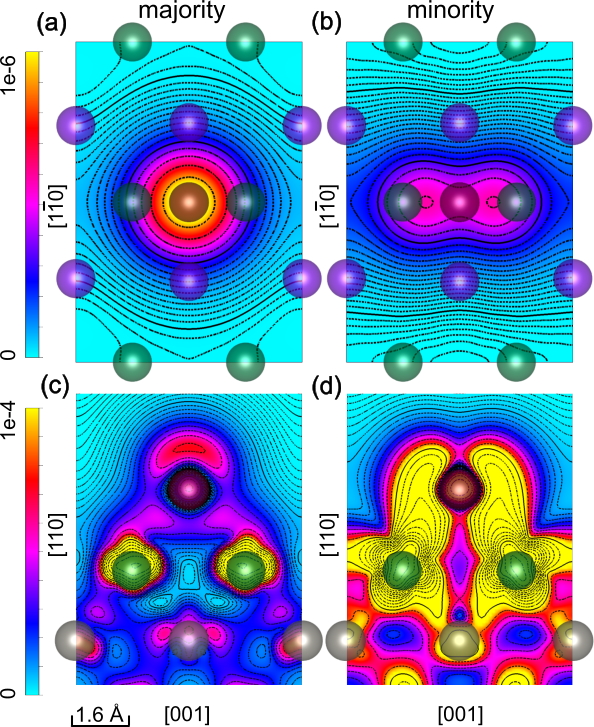}
	\caption{(a, b) Spin-resolved partial charge density plots at 3 {\AA} above the Pb adatom on Mn/W(110) in the energy range [$E_\text{F}-0.065$, $E_\text{F}-0.045$~eV]. (c, d) cross-sectional plots through the Pb adatom parallel to the [001] direction for the charge densities of the top panel. }
	\label{fig:chargedensity_Pb}
\end{figure}

Previously, it has been shown that the spin direction of adsorbed Co adatoms on Mn/W(110) can be detected in spin-polarized STM images at small bias voltages due to the different orbital symmetry
of $d$ states in majority and minority spin channel \cite{Serrate2010}. We find a similar effect for the $p$ orbitals of Pb close to the Fermi energy, $E_F$.
%SH
Figure~\ref{fig:chargedensity_Pb} shows top and cross-sectional spin-resolved partial charge density plots in a small energy window [{\eF}$-$0.065, {\eF}$-$0.045 eV] for Pb adatom adsorbed on Mn/W(110).
A strong interaction between the minority $p_x$ states of the adatom and the minority $d_{z^2}$ orbitals of the neighboring Mn atoms is clearly visible in the cross-sectional plot along the $[001]$ direction shown in Figs.~\ref{fig:chargedensity_Pb}(d). Here, the axes of the $d_{z^2}$ Mn orbitals are distorted pointing towards the Pb atom and a large part of the charge density is concentrated at the interface between adsorbate and substrate. However, such hybridization is less prominent in the majority channel which displays the rotationally symmetric shape of a $p_z$ orbital 
[Fig.~\ref{fig:chargedensity_Pb}(c)].

The partial charge density calculated at a height of 3 {\AA} in the vacuum [Fig.~\ref{fig:chargedensity_Pb}(a-b)] shows that the both spin channels are clearly distinguishable from each other due to the shape of their orbitals. For the majority channel one can clearly observe the $p_z$ states of the adatom in the vacuum. In the minority channel, the double-lobed structure of the  
$p_x$ state protrudes rotationally symmetric states such as $s$, $p_z$ and $d_{z^2}$ orbitals which usually extend further into the vacuum. Similar behavior has been reported previously by Serrate {\etal} for different $d$-states of a Co adatom adsorbed on Mn/W(110)~\cite{Serrate2010}. Hence, we can conclude that in an STM experiment with a magnetic tip it will be possible to identify the spin direction of the Pb adatom by means of the respective orbitals dominating near {\eF} yielding similar effects observed in spin-polarized STM~\cite{Serrate2010,Serrate2016,Haldar2018}.

However, for the Bi adatom the above mentioned feature is not present in the vicinity of $E_F$ which is accessible for STM. In this case, the majority $p_y$ states of Bi are completely covered by the rotationally symmetric orbitals in the vacuum (not shown). Therefore, the orbital shapes for majority and minority states for the charge densities calculated in the vicinity of {\eF} do not differ 
from one another. 

\subsubsection{TAMR of Pb and Bi adatoms on Mn/W(110)}
\label{sub2sec:TAMR_adatom}
\begin{figure}[htb]
	\centering
	\includegraphics[scale=0.45,clip]{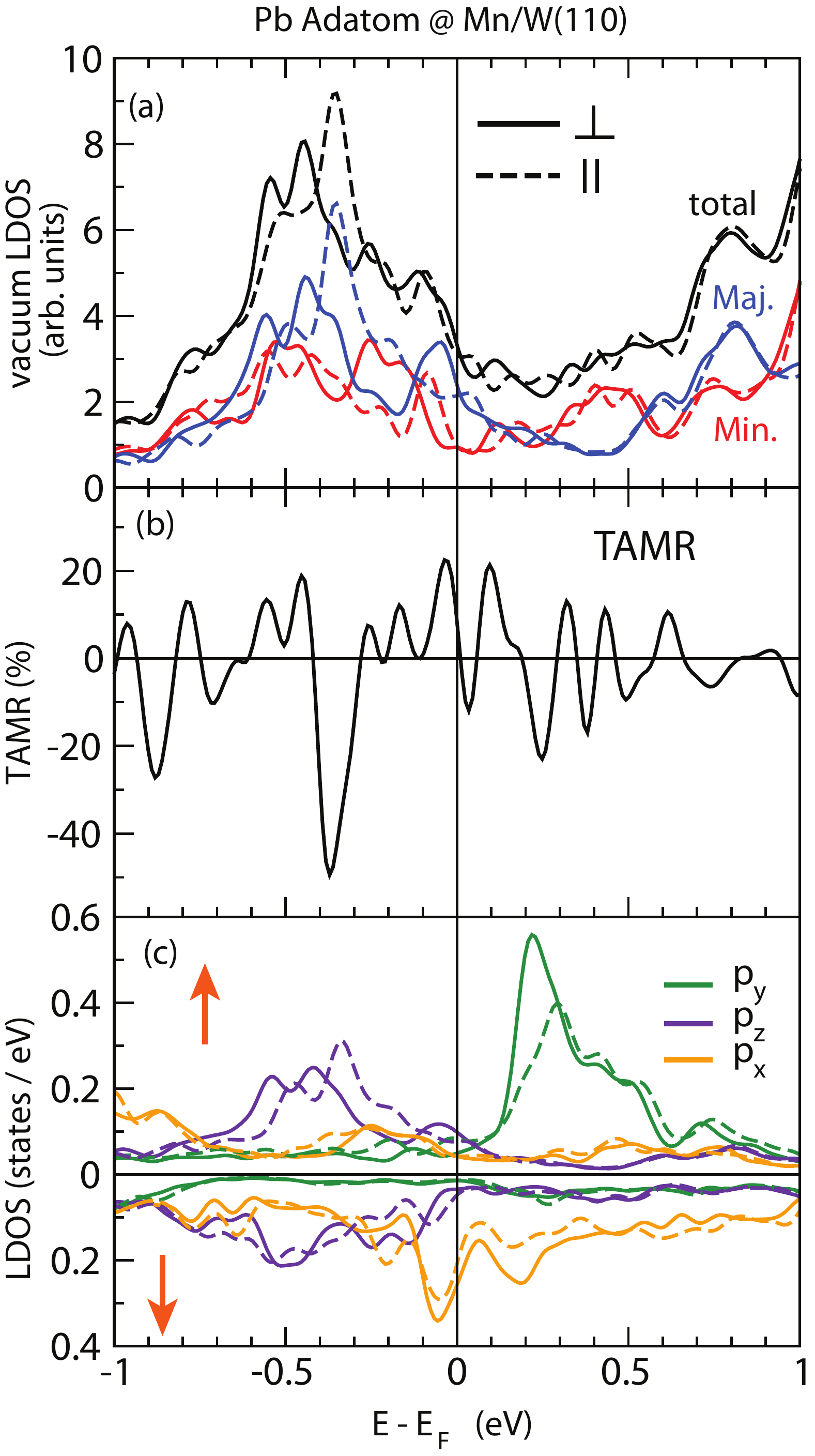}
	\caption{(a) Total (black lines) and spin-resolved (Majority: blue, Minority: red) vacuum LDOS including SOC above the Pb adatom on Mn/W(110) for out-of-plane ($\perp$, solid lines) and in-plane (parallel to the $[1\overline{1}0]$ direction) magnetizations ($\parallel$, dashed lines). (b) TAMR obtained from the spin-averaged vacuum LDOS according to Eq.~(\ref{eq:TAMR}). (c) Orbital decomposition of the LDOS of the Pb adatom in terms of the majority (up) and minority (down) states. Solid (dashed) lines correspond to the magnetization direction perpendicular (parallel) to the surface plane.
	The orange up and down arrow indicates majority and minority spin channels, respectively.}
	\label{fig:PbAdatom}
\end{figure}
\begin{figure}[htb]
	\centering
	\includegraphics[scale=0.45,clip]{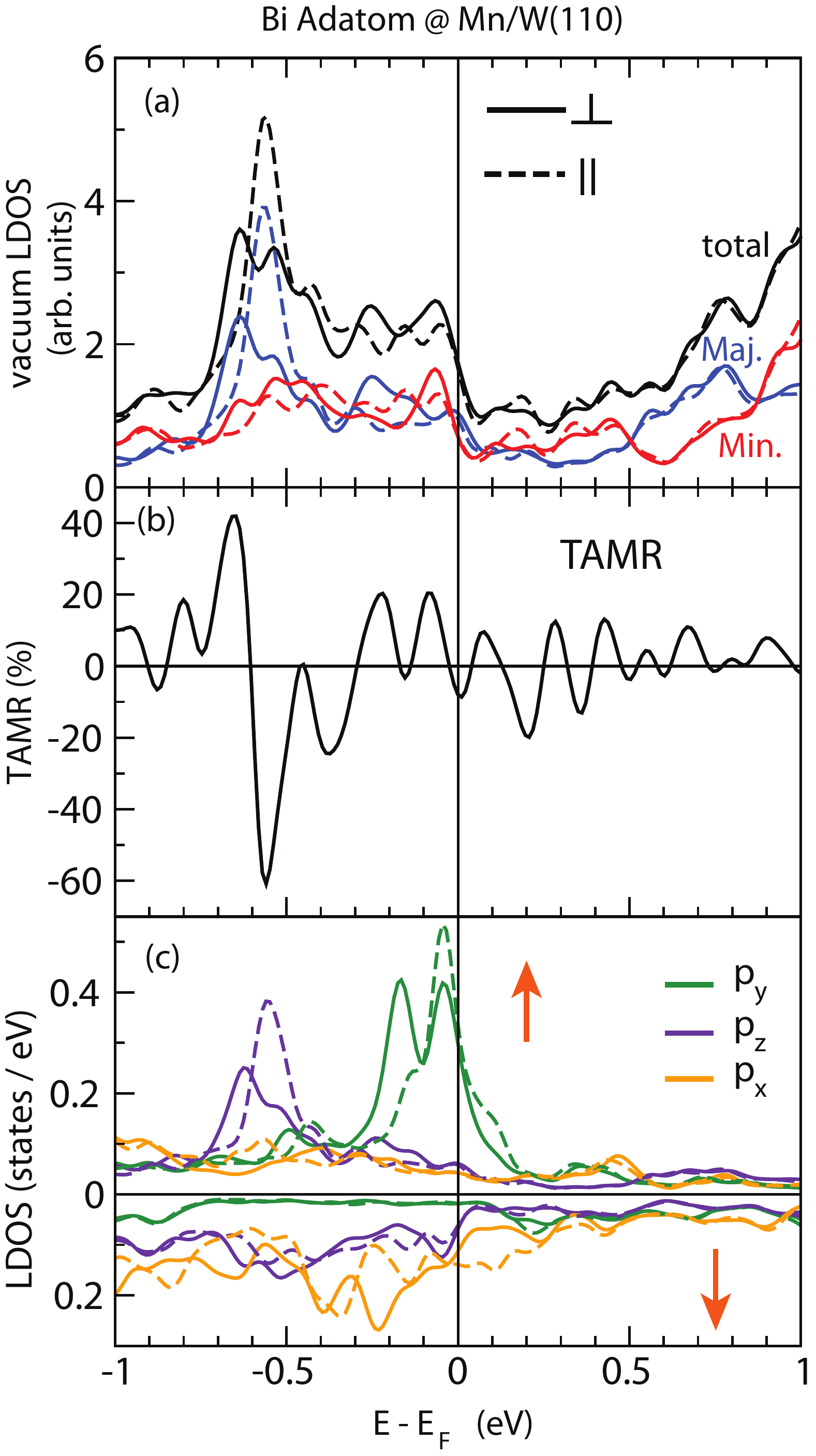}
	\caption{(a) Total (black lines) and spin-resolved (Majority: blue, Minority: red) vacuum LDOS including SOC
  above the Bi adatom on Mn/W(110) for out-of-plane ($\perp$, solid lines) and in-plane (parallel to the $[1\overline{1}0]$ direction) magnetizations ($\parallel$, dashed lines). (b) TAMR obtained from the spin-averaged vacuum LDOS according to Eq.~(\ref{eq:TAMR}). (c) Orbital decomposition of the LDOS of the Bi adatom in terms of the majority (up) and minority (down) states. Solid (dashed) lines correspond to the magnetization direction perpendicular (parallel) to the surface plane.
	The orange up and down arrow indicates majority and minority spin channels, respectively.}
	\label{fig:BiAdatom}
\end{figure}

In this section we will focus on the description of the electronic structure of $6p$ adatoms adsorbed on the Mn monolayer of W(110). Especially the anisotropy of the LDOS due to SOC and the subsequent TAMR effect will be discussed in detail.

Fig.~\ref{fig:PbAdatom}(a) shows both the total (spin-averaged) and spin-resolved vacuum LDOS above the Pb adatom 
%SH
-- in an energy range around $E_F$ typically accessible to STM --
calculated for the two magnetization directions including SOC:
(i) perpendicular to the surface (out-of-plane) denoted as $n_\perp(z, \epsilon)$ and parallel to the $[1\overline{1}0]$ direction (in-plane) denoted as $n_\parallel(z, \epsilon)$.
%The peaks in the vacuum LDOS are dominated by $p_z$ states which decays more slowly perpendicular to the surface. 
Differences between both magnetization components are clearly discernible in the energy range below the Fermi level ({\eF}). The most significant feature is located at $-0.37$~eV in $n_\parallel(z, \epsilon)$ and corresponds to a peak of majority $p_z$ states being split and shifted towards lower energies as the magnetization rotates from the film plane ($\parallel$) to the perpendicular ($\perp$) direction of the surface. The same effect, although much less prominent, is also  visible for the minority states. This behavior leads to a maximum value in the TAMR of $-49$\% (see Fig.~\ref{fig:PbAdatom}(b)). Around {\eF} this effect is considerably smaller and of opposite sign with TAMR values up to +22\%. 

Similar observations can be seen in the vacuum LDOS of the Bi adatom on Mn/W(110) shown in Fig.~\ref{fig:BiAdatom}(a). Here, the dominant peak of majority $p_z$ states which splits likewise upon rotation of the magnetization direction is shifted by 0.2~eV towards lower energies compared to Pb. Linked to this state, the
value of the TAMR 
first takes a local maximum of +42\% at $-0.66$~eV before dropping abruptly to a minimum of $-61$\% at $-0.57$~eV below {\eF}. Similar to Pb, differences concerning $n_\perp$ and  $n_\parallel$ for the minority channel are small in this energy range and the main part of the TAMR originates from majority states. In contrast, states with minority character are causing a modest TAMR of +20\% just below  {\eF}. For both adatoms the anisotropy of the vacuum LDOS shows only little magnetization-direction dependent differences in the unoccupied regions and giant values in the TAMR effect are restricted to areas below {\eF}.

A closer look at the orbitally resolved LDOS of the adatoms in Fig.~\ref{fig:PbAdatom}(c) and Fig.~\ref{fig:BiAdatom}(c) reveals that the above-mentioned changes between both magnetization components in the vacuum can be attributed to $p_z$ states of the adatoms which are mostly below {\eF}.
The curves in the vacuum almost coincide with the ones for states of this character calculated directly at the respective adatom.
As the $p_z$ states are oriented along the surface normal,
they preponderate in the vacuum compared to the $p_x$ and $p_y$ states. In contrast, the prominent peak of majority $p_y$ states dominating the LDOS in the vicinity of {\eF} of both Pb and Bi is not visible in the vacuum LDOS because 
they are 
aligned parallel to the film plane. 
The shift of this peak, from a position of 0.2 eV above {\eF} for Pb, towards occupied regions for Bi can be explained by the increasing number of electrons in the $p$ shell. On the other hand, the shift of the majority states with $p_z$ character which are identified to generate the large anisotropy of the vacuum LDOS and hence the shift of the position of the maximum TAMR can be ascribed to the different strength of the attractive potential acting between valence electrons and nucleus. Due to the larger nuclear charge these potentials lead to a stronger binding of the $p_z$ states to the nucleus for Bi. Further reasons for the majority $p_z$ states of the Bi adatom being shifted towards lower energies is the higher spin polarization compared to Pb as well as the smaller distance from its nearest neighbor Mn atom in the Mn monolayer. Hereby the orbital overlap increases resulting in a larger splitting of the states.

\subsubsection{Modeling of the TAMR}
In order to explain the large TAMR found for $6p$ adatoms adsorbed on Mn/W(110), we revert to the Hamiltonian of SOC mentioned in the introduction. As shown in Ref.~\cite{Abate1965}, the SOC operator can be written as a matrix in the following way: 
\begin{equation}
 \mathcal{H}_{SOC}= \frac{\xi}{2}
 \begin{pmatrix}
    M&N\\
    -N^{*}&M^{*}
\end{pmatrix}\; .
\end{equation}
Here, the diagonal matrices $M$ describe the coupling of two states with equal spin direction, whereas the secondary diagonal matrices $N$ denote the interaction of states with different spin character via SOC. Both can be calculated for an arbitrary orientation of the spin quantization axis by applying ladder operators of spin and angular momentum to linear combinations of complex spherical harmonics which represent both $p$ and $d$ orbitals. This approach yields the matrix element describing a spin-orbit induced hybridization between states with $p_z$ and $p_x$ symmetry in the same spin channel as~\cite{Schoeneberg2016Diss}:
\begin{equation}
    \langle \uparrow,p_z|\mathcal{H}_{SOC}|p_x, \uparrow \rangle= i\sin\theta\sin\phi \;, 
    \label{eq:H_SOC_up_up}
\end{equation}
and the element for coupling states of the same symmetry, but with opposite spin direction as
\begin{equation}
    \langle \uparrow,p_z|\mathcal{H}_{SOC}|p_x, \downarrow \rangle=\cos\phi+i\sin\phi\cos\theta \;. 
    \label{eq:H_SOC_up_down}
\end{equation}

In the first case (cf. Eq.~(\ref{eq:H_SOC_up_up})) the matrix element vanishes for the perpendicular magnetization direction ($\phi$=0$^{\circ}$, $\theta$=0$^{\circ}$) and becomes maximal for its magnetization pointing along the $[1\overline{1}0]$ direction ($\phi$=90$^{\circ}$, $\theta$=90$^{\circ}$), i.e. we expect a mixing of the two states only for a spin-quantization axis chosen along the film plane. The reverse is true if both states have opposite spin direction (cf. Eq.~(\ref{eq:H_SOC_up_down})). Evaluating the matrix elements given in Ref.~\cite{Schoeneberg2016Diss} for a potential hybridization mediated by SOC for states with $p_z$ and $p_y$ character shows that such interaction can not be realized on the Mn/W(110) surface for the two above mentioned magnetization directions, which are possible on the substrate due to the spin spiral ground state. 
%Both for the same as well as for the opposite spin channel they vanish \cite{Gutzeit2018}.
For this reason the discussion concerning the anisotropy of the vacuum LDOS is restricted to $p_x$ and $p_z$ states for both $6p$ adatoms and dimers in this paper.

Applying the above considerations first to the case of a Bi adatom on Mn/W(110) [Fig.~\ref{fig:BiAdatom}], one can explain the maximum value of the TAMR at $-0.57$~eV below {\eF} by a magnetization-direction dependent mixing of $p_x$ and $p_z$ orbitals of opposite spin channels. At this energy the prominent peak of majority $p_z$ states whose in-plane magnetization component resembles a single peak is split and shifted towards lower energies upon rotation of the spin-quantization axis (see Fig.~\ref{fig:BiAdatom}(c)).  According to the matrix elements, this behavior hints at a SOC-mediated hybridization with a minority $p_x$ state which can be found at $-0.82$~eV.

The TAMR of the Pb adatom can also be understood based on the matrix elements of $H_{\rm SOC}$. E.g.~the vacuum LDOS of the minority spin channel [Fig.~\ref{fig:PbAdatom}(a)] just
below $E_F$ is reduced upon rotating the magnetization direction from in-plane to out-of-plane. This is due to mixing by SOC in the minority spin channel [Fig.~\ref{fig:PbAdatom}(c)] 
between a $p_z$ state located at $-0.12$~eV and a peak at $-0.05$~eV of $p_x$ orbital character. For an in-plane magnetization direction, which allows mixing within the same spin channel 
by SOC according to Eq.~(\ref{eq:H_SOC_up_up}), the $p_x$ minority state peak at $-0.05$~eV splits into two peaks which coincide with the positions of two minority states $p_z$ peaks. 
%is reduced in height while the $p_z$ peak splits and one peak is shifted in the direction of {\eF}. 
This creates a large negative TAMR within the minority spin channel of $-56$\% (not shown here).
However, the TAMR is obtained from the total, spin-averaged LDOS. Just below the Fermi energy it is positive with a value of $+22$\% due to a majority $p_z$ peak whose height is
reduced due to SOC for an in-plane magnetization [Fig.~\ref{fig:PbAdatom}(c)]. 
%SH
The maximum TAMR effect of the Pb adatom of $-49$\% occurs at $0.37$ eV below {\eF}. It originates from the majority spin channel [Fig.~\ref{fig:PbAdatom}(a)] and it is due to
the splitting of a majority $p_z$ state as can be seen from the orbital decomposition at the Pb atom [Fig.~\ref{fig:PbAdatom}(c)]. Since the mixing occurs for a magnetization
direction perpendicular to the surface it can be explained by a SOC induced mixing with $p_x$ states of the opposite spin channel according to Eq.~(\ref{eq:H_SOC_up_down}).
While changes in the minority $p_x$ LDOS can be noted within the relevant energy interval it is not possible to unambiguously propose a single peak which is responsible for
the mixing. As will be discussed in detail for the Pb dimers at the end of this manuscript, there is also an impact of the Mn $3d$ states which are also subject to SOC and
with which the Pb $p$ states are hybridizing.
%At this point the question remains which states are involved in the maximum TAMR of $-49$\% at $-0.37$ eV below {\eF};  the overlap of the Pb $p$ orbitals and their delocalization makes it difficult to allocate a hybridization partner for the prominent majority $p_z$ state according to the mentioned matrix elements. This observation thus indicates that a simple model considering only two atomic states might not be sufficient in this case to describe their interactions accordingly.     

\subsection{Pb and Bi dimers on Mn/W(110)}
\label{subsec:dimers}
\subsubsection{Structural and magnetic properties}
\label{sub2sec:str_mag_dimers}
\begin{table}[htb]
	\centering
	\caption{The dimer bond lengths $d$ (in \AA) and the individual magnetic moments (in {\uB}) of the adsorbed Pb and Bi dimers on the Mn monolayer on W(110). For comparison, the bond lengths and the magnetic moments of the free dimers (calculated with SOC) are given.}
	\label{tab:dimer_d_m}
	\begin{ruledtabular}
	\begin{tabular}{c c c c c}
		\multirow{2}{*}{Orientation} & \multicolumn{2}{c}{Pb} & \multicolumn{2}{c}{Bi} \\ %\cline{2-5} 
		& $d$ & {\uB} & $d$ & {\uB} \\
		\colrule
		$[001]$ & 3.23 & +0.08 & 3.22 & +0.12 \\ 
		$[1\overline{1}0]$ & 3.35 & +0.18 & 3.93 & +0.02 \\ 
		$[1\overline{1}1]$ & 3.11 & $\pm$0.02 & 3.12 & $\pm$0.02 \\ 
		Free & 2.96 & 0.67 & 2.68 & 0.0 \\
	\end{tabular} 
	\end{ruledtabular}
\end{table}
Since the Pb and Bi adatoms adsorb in the hollow-site position of the Mn layer, the dimers can be oriented along the $[001]$, $[1\overline{1}0]$, and $[1\overline{1}1]$ directions (see Fig.~\ref{fig:geom}(c)). The relaxed dimer bond lengths and the magnetic moments for the three orientations along with the values for free dimers are given in Table~\ref{tab:dimer_d_m}. The dimer bond lengths increase after the adsorption due to structural relaxation from the bond length values of free dimers. For Pb dimers an increase of $\approx$ 10\% in bond length can be observed. For Bi dimers, a larger increase of $\approx$ 20\% in bond length has been observed except along the $[1\overline{1}0]$ orientation. In this case, we find an increase of bond length values by $\approx$ 45 \%. 

In Pb dimers, the individual atoms carry small induced magnetic moments for all three orientations due to the hybridization with the Mn monolayer. Among the three orientations, the largest individual magnetic moment of +0.18 {\uB} is observed for the $[1\overline{1}0]$ orientation. These induced moments for Pb dimers are in contrast with the single atom adsorption where Pb remains nonmagnetic. Similar to the single adatom adsorption, Bi dimers also pick up small induced magnetic moment for all orientations with the largest value of +0.12 {\uB} along the $[001]$ direction. Similar to the single adatom adsorption, reduction of magnetic moments for both nn and nnn Mn adatoms have been observed here as well.

\subsubsection{Electronic properties}
\begin{figure*}[htbp]
	\centering
	\includegraphics[scale=0.5,clip]{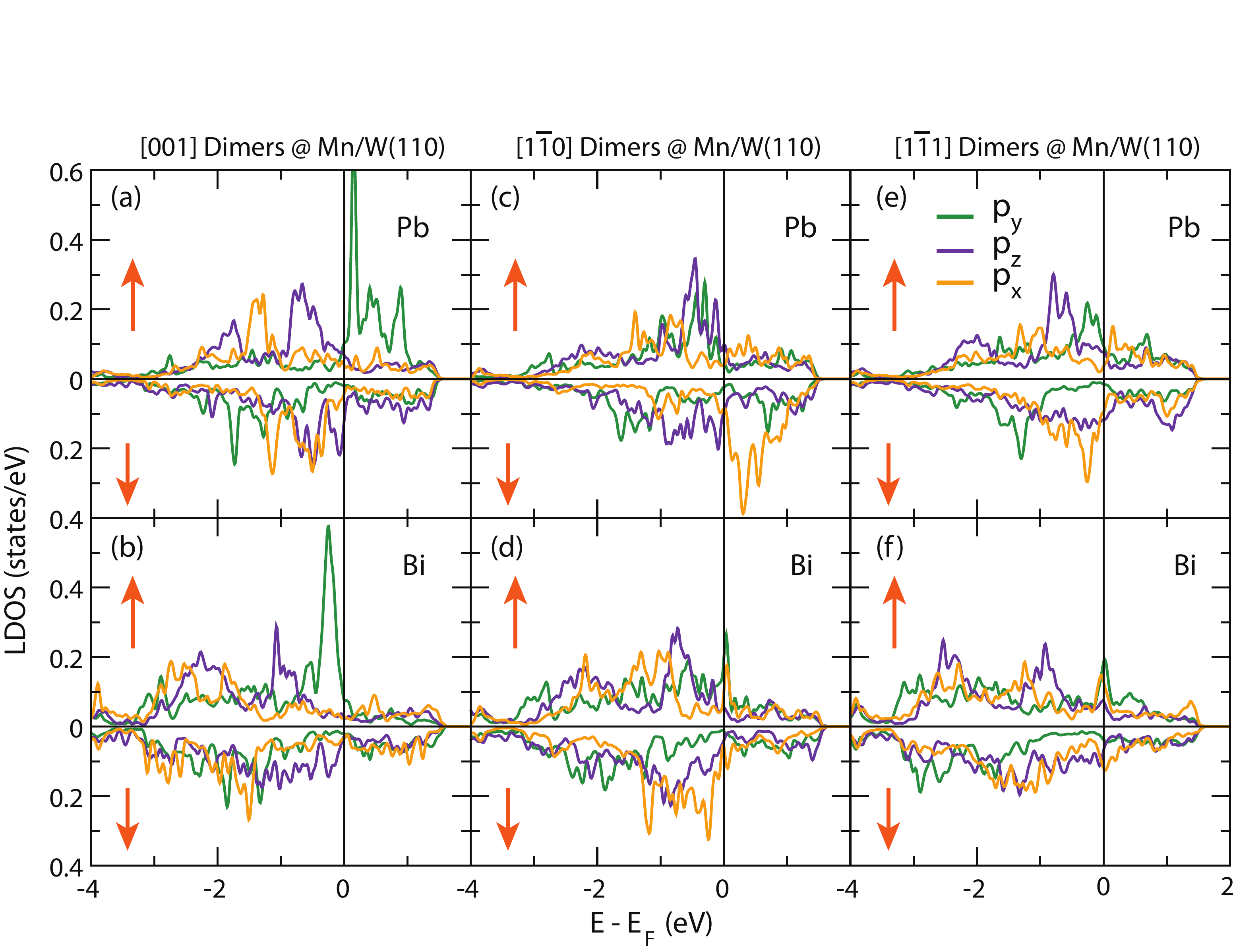}
	\caption{(a) Scalar relativistic spin and $m_l$ resolved LDOS of the Pb and Bi dimers adsorbed on Mn/W(110) for (a, b) the $[001]$ orientation, (c, d) $[1\overline{1}0]$ orientation, and (e, f) $[1\overline{1}1]$ orientation. The orange up and down arrow indicates majority and minority spin channels, respectively.}
	\label{fig:Pb_Bi_dimer_dos_full}
\end{figure*}
We proceed by describing and comparing the electronic structure of the dimers with those presented for the single adatoms before explaining the anisotropy of the LDOS. It should be pointed out here that the notation of the $p$ orbitals of the dimers refers to the global coordinate axes of the Mn/W(110) surface as shown in Fig.~\ref{fig:geom}, i.e. no local system for the adsorbates  rotated for different orientations has been used. Therefore, the $p_x$ and $p_y$ orbitals of both [001] and $[1\overline{1}0]$ dimers are aligned along the [001] and $[1\overline{1}0]$ direction, respectively. As a result, the orbitals responsible for the covalent bond are changing.

Fig. \ref{fig:Pb_Bi_dimer_dos_full} shows the $m_l$ decomposed $p$ states of both Pb and Bi dimers on a large energy scale around {\eF}. We will exemplify the differences in the LDOS compared to the single adatoms by means of the Bi dimers; similar observations can be made for the respective Pb adsorbates. Compared with the Bi adatom (cf. Fig.~\ref{fig:spinpol_adatom}(f)), the Bi dimer along the $[001]$ orientation exhibits the largest modifications in its $p_x$ orbitals which are responsible for the covalent bond in this case hereby forming $\sigma$ orbitals (cf. Fig.~\ref{fig:Pb_Bi_dimer_dos_full}(b)). This becomes most evident in the minority channel just below {\eF} where the corresponding states of the single adatom are shifted by 1 eV to the left due to the orbital overlap of the two atoms of the dimer. In contrast, the $p_y$ and $p_z$ states only show minor differences compared to the Bi adatom; especially the large peak of majority $p_y$ states dominating close to {\eF} of the adatom is also found for the $[001]$ dimer. 

Owing to the large distance of 3.93 {\AA} between the two Bi atoms of the $[1\overline{1}0]$ dimer the overlap of their orbitals is small resulting in similar features as for the adatom (cf. Fig.~\ref{fig:Pb_Bi_dimer_dos_full}(d)). The main change in its $m_l$ resolved LDOS is the disappearance of the dominant majority $p_y$ peak at {\eF} which can be attributed to the fact that these orbitals are forming $\sigma$ bonds in this case. For the Bi $[1\overline{1}1]$ dimer, differences in the $m_l$ decomposed LDOS (cf. Fig.~\ref{fig:Pb_Bi_dimer_dos_full}(f)) are clearly discernible compared to the single adatom (cf. Fig.~\ref{fig:spinpol_adatom}(f)). This observation can partly be ascribed to the small bond length of 3.12 {\AA} and hence a large orbital overlap and partly to the combination of $p_x$ and $p_y$ states forming $\sigma$ bonds. Both orbitals are tilted with respect to the global coordinate axes of the substrate leading to a covalent bond of a mixture of the two states with different symmetry. The $p_z$ states which are crucial for STM are relatively weakly affected for all dimers.   

\subsubsection{TAMR effect of the Pb dimers on Mn/W(110)}
\label{sub2sec:TAMR_dimers_Pb}
\begin{figure*}[htbp]
	\centering
	\includegraphics[scale=0.37,clip]{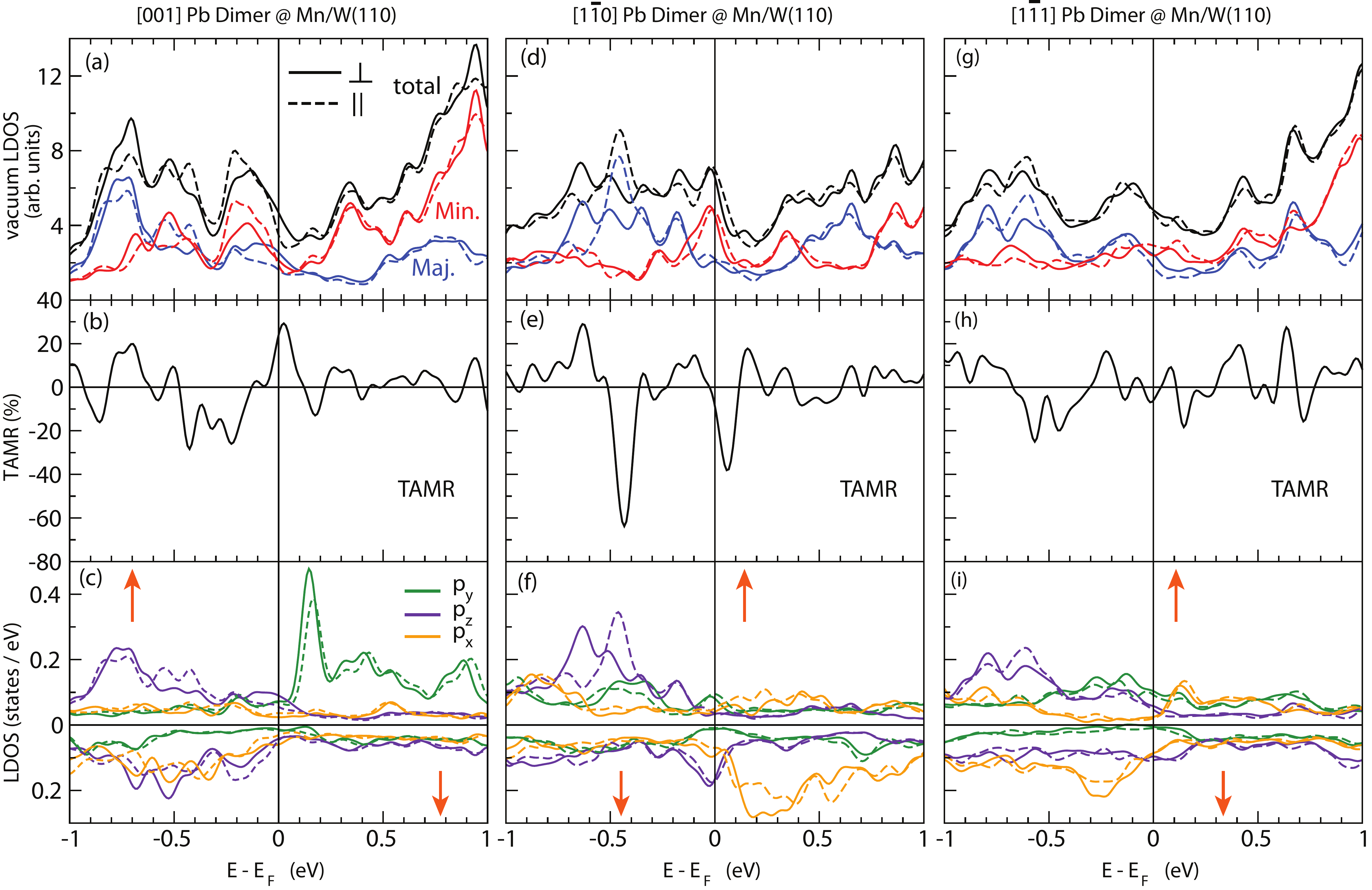}
	\caption{(a, d, g)Total (black lines) and spin-resolved (Majority: blue, Minority: red) vacuum LDOS  above the Pb dimer on Mn/W(110) for out-of-plane ($\perp$, solid lines) and in-plane (parallel to the $[1\overline{1}0]$ direction) magnetizations ($\parallel$, dashed lines). (b, e, h) TAMR obtained from the spin-averaged vacuum LDOS according to Eq.~(\ref{eq:TAMR}). (c, f, i) Orbital decomposition of the LDOS of the Pb dimers in terms of the majority (up) and minority (down) states. Solid (dashed) lines correspond to the magnetization direction perpendicular (parallel) to the surface plane. The orange up and down arrow indicates majority and minority spin channels, respectively.
	}
	\label{fig:Pb_dimer_dos}
\end{figure*}
In the following section we study the TAMR of Pb dimers on Mn/W(110) for the three different dimer orientation discussed before. 

In Fig.~\ref{fig:Pb_dimer_dos}(a) both the total and spin-resolved components of the vacuum LDOS of the Pb dimer oriented along the [001] direction are plotted for the two different magnetization
directions which can occur due to the spin spiral states of the Mn monolayer on W(110). 
%SH
Note, that the in-plane magnetization direction is perpendicular to the dimer axis in this case.
Compared to the respective adatom (cf. Fig.~\ref{fig:PbAdatom}), only small differences between $n_{\perp}$ and $n_{\parallel}$ can be observed in the occupied regions below {\eF}. The most striking features occur now at energies 
$-0.7$ eV,
$-0.42$~eV,
$-0.2$~eV, and  
$0.05$~eV leading to maximum values in the TAMR of $\pm 28$\% (see Fig.~\ref{fig:Pb_dimer_dos}(b)). 
Unlike the Pb adatom, the anisotropy of the vacuum LDOS takes another local maximum of +29\% just above {\eF}.
As one can see from the spin-resolved curves 
%SH
in Fig.~\ref{fig:Pb_dimer_dos}(a), the TAMR at $-0.7$~eV stems from a modification of both spin channels upon rotation of the magnetization direction, whereas at {\eF} only majority states contribute whose perpendicular magnetization components are clearly enhanced compared to the parallel ones. Similar to the Pb adatom, differences between $n_{\perp}$ and $n_{\parallel}$ in the unoccupied regions are barely noticeable for both spin directions of the [001] dimer. 

The orbitally decomposed LDOS of this dimer plotted in Fig.~\ref{fig:Pb_dimer_dos}(c), shows further similarities with the adatom. It is dominated by a prominent peak of majority $p_y$ states at 0.15 eV above {\eF} which is not reflected in the vacuum LDOS and exhibits discernible changes in the $p_z$ states with respect to both magnetization directions below {\eF}. The $p_z$ states  predominate the vacuum LDOS due to their double-lobed orbitals pointing along the surface normal. However, they experience a small shift towards lower energies as well as a splitting which is a consequence of the interaction between both Pb atoms composing the dimer.  

Compared to the $[001]$ Pb dimer, the anisotropy of the vacuum LDOS is much larger for the dimer oriented along the $[1\overline{1}0]$ direction representing the propagation direction of the spin spiral on Mn/W(110) (see Fig.~\ref{fig:Pb_dimer_dos}(d)-(e)). 
%SH
For this dimer orientation the in-plane magnetization direction is along the dimer axis (cf.~Fig.~\ref{fig:geom}).
As for the single adatom, the appearance of the LDOS in the vacuum below {\eF} is characterized  by magnetization-direction dependent differences of the majority $p_z$ states where the main contribution comes from a dominant peak of the in-plane magnetized dimer at $-0.45$ eV below {\eF}. Being split multiple times upon reorientation of the spin-quantization axis, it creates a steep descent in the TAMR up to $-64$\% thereby even exceeding the maximum value of the Pb adatom by 15\%. Just above {\eF}, the anisotropy of the vacuum LDOS takes a local minimum of $-38$\%  which is due to the shift of a minority $p_z$ state as the magnetization rotates from the perpendicular direction to the film plane along the $[1\overline{1}0]$ direction. The substantial similarity of the LDOS of this dimer in the $p_z$ orbitals compared to the single adatom (cf.~Fig.~\ref{fig:PbAdatom}) can be explained by means of the relatively large distance of both Pb atoms (see Table \ref{tab:dimer_d_m}). If they are further apart, their interaction, i.e. the overlap of their orbitals, will be small thereby causing a similar behavior as for a single atom (cf. Fig.~\ref{fig:PbAdatom}(c) and Fig.~\ref{fig:Pb_dimer_dos}(f)). The $p_y$ orbitals, on the other hand, which are forming $\sigma$ bonds in the case of the $[1\overline{1}0]$ dimer are expected to show more remarkable differences in comparison with the adatom.
This becomes mostly evident in the unoccupied regions where the prominent peak of majority $p_y$ states is completely absent as seen in Fig.~~\ref{fig:Pb_dimer_dos}(f) with respect to the adatom (cf. Fig.~\ref{fig:PbAdatom}(c)).    

\begin{figure*}[hbt!]
\centering
\includegraphics[scale=0.37,clip]{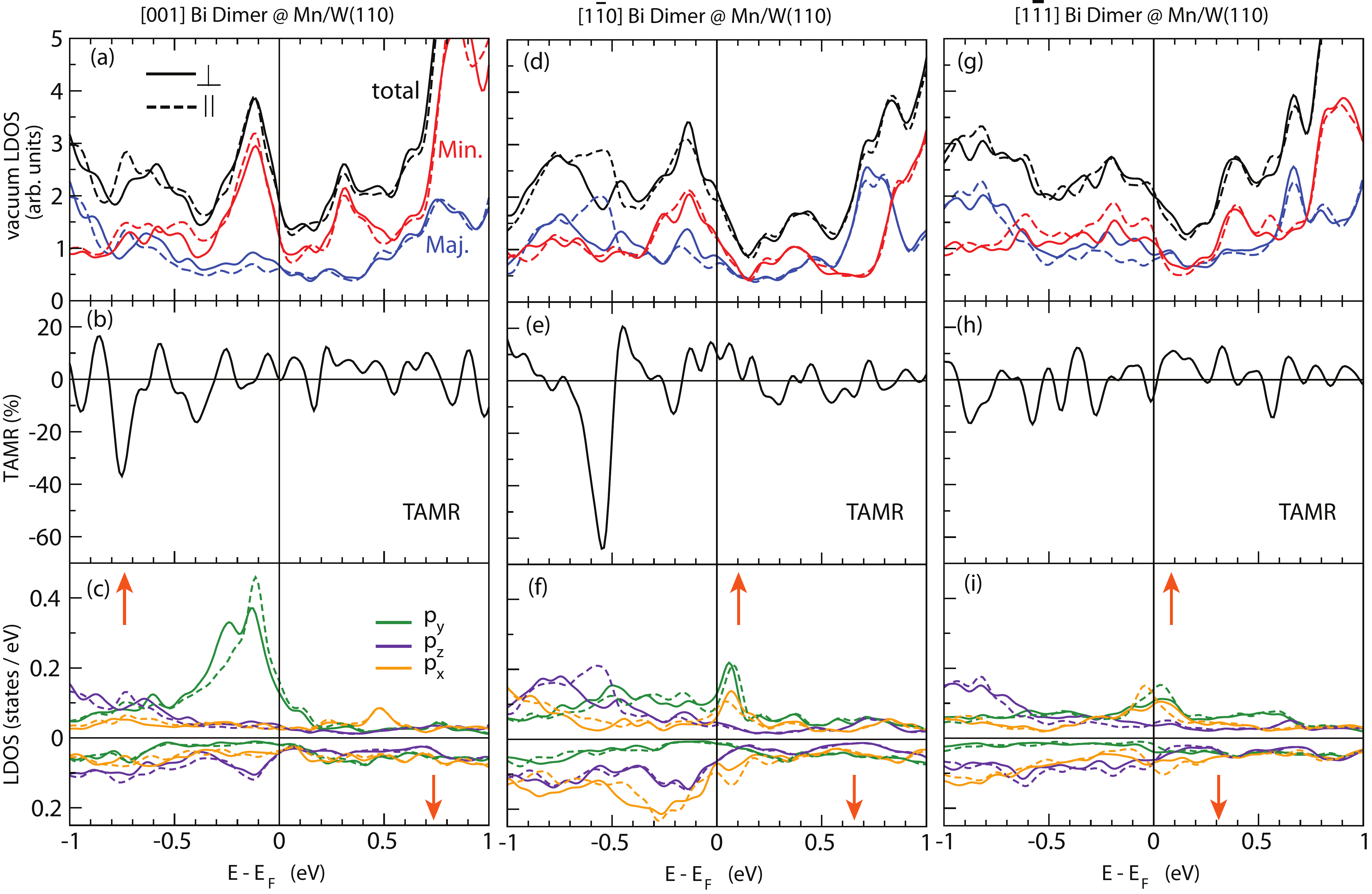}
\caption{(a, d, g)Total (black lines) and spin-resolved (Majority: blue, Minority: red) vacuum LDOS  above the Bi dimer on Mn/W(110) for out-of-plane ($\perp$, solid lines) and in-plane (parallel to the $[1\overline{1}0]$ direction) magnetizations ($\parallel$, dashed lines). (b, e, h) TAMR obtained from the spin-averaged vacuum LDOS according to Eq.~(\ref{eq:TAMR}). (c, f, i) Orbital decomposition of the LDOS of the Pb dimers in terms of the majority (up) and minority (down) states. Solid (dashed) lines correspond to the magnetization direction perpendicular (parallel) to the surface plane. 
The orange up and down arrow indicates majority and minority spin channels, respectively.}
\label{fig:Bi_dimer_dos}
\end{figure*}

The electronic structure of the Pb dimer along $[1\overline{1}1]$ direction is shown in Fig.~\ref{fig:Pb_dimer_dos}(g)-(i) for both magnetization directions. As for the [001] Pb dimer, only minor differences between $n_{\perp}$ and $n_{\parallel}$ are visible in the vacuum LDOS above the $[1\overline{1}1]$ dimer along the diagonal of the unit cell. At $-0.6$ eV below {\eF},
a peak of majority states for parallel magnetization direction is enhanced compared to the perpendicular magnetization direction resulting in a negative TAMR of $-25$\%. The vacuum LDOS of the unoccupied spectrum on the other hand is mainly characterized by magnetization-direction dependent changes of the minority states causing maximum values of +28\% in the TAMR at +0.62 eV. 

The magnitude of TAMR is much smaller for $[1\overline{1}1]$ dimer orientation as compared to the $[1\overline{1}0]$, which is oriented along the natural magnetization direction of the Mn/W(110) substrate. However, it is very similar as for the dimer oriented along $[001]$, which is perpendicular to the magnetization of the surface. We will discuss these behavior in more detail in Section \ref{TAMR_origin_Dimers}.  

\subsubsection{TAMR of Bi dimers on Mn/W(110)}
\label{sub2sec:TAMR_dimers_Bi}
Similar changes of the TAMR with the dimer orientation can be observed for Bi dimers on Mn/W(110). Considering first the electronic structure of the [001] Bi dimer which is shown in Fig.~\ref{fig:Bi_dimer_dos}(a)-(c), one notices that as for the [001] Pb dimer the curves for both magnetization directions of the spin-averaged vacuum LDOS do not differ significantly from each other. The largest differences are now located in the energy range between $-0.9$ and $-0.3$ eV leading to a local maximum in the TAMR of $-37$\% at $-0.75$ eV. The negative TAMR value is much smaller than for the Bi adatom (cf. Fig.~\ref{fig:BiAdatom}(b)). 
%SH because the in-plane magnetization components of both spin channels are enhanced compared to the out-of-plane magnetization. 
Consistent with the single Bi adatom, the energy range just below {\eF} is dominated by a large peak of minority states with $p_z$ character (see Fig.~\ref{fig:Bi_dimer_dos}(c)), whereas the prominent peak of majority $p_y$ orbitals is absent in the vacuum LDOS due to its orientation within the film plane. 

In contrast to the anisotropy of the vacuum LDOS of the corresponding Pb dimer, changes between $n_{\perp}$ and $n_{\parallel}$ vanish in the case of the [001] Bi dimer directly at {\eF}. A closer look at the orbitally resolved LDOS reveals that the $p_z$ states which dominate the LDOS above the surface are affected by the mutual interaction of both Bi atoms composing the dimer. In comparison with the single Bi  adatom, they are shifted towards lower energies and experience a larger splitting which is mostly apparent in the majority channel. The dominant peak of majority $p_z$ states causing the large TAMR of $-61$\% at $-0.57$ eV for the single Bi adatom (cf. Fig.~\ref{fig:BiAdatom}(b)) is therefore located outside of the presented energy range for the dimers. However, owing to the interaction of the orbitals it is split as well and less pronounced than for the single Bi adatom (not shown).

The hybridization of majority $p_z$ orbitals is less prominent in the case of a Bi dimer oriented along the $[1\overline{1}0]$ direction (Fig.~\ref{fig:Bi_dimer_dos}(d)-(f)) 
of the Mn/W(110) surface due to the large distance of nearly 4 {\AA} between both Bi atoms. As one can see from its orbital decomposition in Fig.~\ref{fig:Bi_dimer_dos}(f), $p_z$ states
move closer to {\eF} showing similar characteristics 
as in the case of the single Bi adatom (cf. Fig.~\ref{fig:BiAdatom}). At $E_{\rm F}-0.55$ eV  a dominant peak of majority $p_z$ states for the in-plane magnetization orientation becomes visible which both decreases in height and shifts towards lower energies upon rotation of the spin-quantization axis. This is also the largest observable change between $n_{\perp}$ and $n_{\parallel}$ of the total (spin-averaged) vacuum LDOS for the $[1\overline{1}0]$ Bi dimer that is plotted in Fig.~\ref{fig:Bi_dimer_dos}(d). The just mentioned magnetization-direction dependent changes in the majority $p_z$ states thereby correspond to a huge TAMR effect of $-64$\% at $-0.55$ eV. Hence, the anisotropy of the LDOS is in the same order of magnitude as for the single Bi adatom and takes its largest value at the same energetic position as well. The TAMR is considerably smaller for the rest of the presented energy range, especially at {\eF} where differences for parallel and perpendicular magnetizations of both spin channels only create a modest TAMR of approximately 15\%.  

As observed for the corresponding Pb dimer, only minor differences between both magnetization directions occur in the spin-averaged vacuum LDOS of the Bi dimer placed along the diagonal of the unit cell, i.e. the $[1\overline{1}1]$ direction. Whereas the spin-resolved curves are indeed clearly characterized by changes upon rotation of the magnetization in the occupied regions (see Fig.~\ref{fig:Bi_dimer_dos}(g)), changes in the sum of both spin channels only lead to small values in the TAMR ranging from $-16$\% at $-0.8$ eV up to $-10$\% at {\eF}. As shown in the orbital decomposed LDOS of the dimer in Fig.~\ref{fig:Bi_dimer_dos}(i), the main part of the anisotropy is created by the $p_z$ orbitals. Additionally, this dimer orientation exhibits the smallest TAMR of all studied Bi dimer geometries and consistent with the previously presented results for $[1\overline{1}1]$  Pb dimers on Mn/W(110). 

\subsubsection{Origin of the TAMR for $6p$ dimers on Mn/W(110)}
\label{TAMR_origin_Dimers}
The variation of the TAMR  magnitude depending on the orientation of both Pb and Bi dimers can partially be explained by means of a physical model considering two atomic states coupled 
via SOC proposed in Ref.~\onlinecite{Neel2013}.
For a possible SOC induced hybridization of the dimer $p$ states, we refer to the matrix elements presented in section~\ref{sub2sec:TAMR_adatom}. The orbitals are defined with respect to the global coordination axes of the unit cell.

Within this simplified model, our expectations match quite well with the $6p$ dimers oriented along the $[1\overline{1}0]$ direction showing the largest TAMR of all studied configurations. However, the SOC induced hybridization is not so clearly 
visible between their $p_z$ and $p_x$ states forming $\pi_z$ and $\pi_x$ molecular orbitals, respectively. The reduction of the TAMR effect for the $[1\overline{1}1]$ $6p$ dimers can be understood as well using the simplified model of two atomic states with differing orbital symmetry. For the case of a dimer orientation along the diagonal of the supercell the magnetization direction of the substrate is rotated with respect to the bonding axis of the atoms leading to a reduction of the respective matrix elements.  
For instance the hybridization between the $p$ states, $ \langle \uparrow,p_z|\mathcal{H}_{SOC}|p_x, \uparrow \rangle \propto \sin\theta\sin\phi$, is reduced for an azimuth angle of  45$^{\circ}$. Since changes in the LDOS scale with the square of the matrix elements~\cite{Bode2002},  
for both $6p$ dimers, the TAMR in $[1\overline{1}1]$ orientation is diminished by a factor of 4 compared to the $[1\overline{1}0]$ dimer ($\sim$ 15\% vs. $\sim$ 60\%; cf. Fig.~\ref{fig:Pb_dimer_dos}(e), Fig.~\ref{fig:Pb_dimer_dos}(h) and Fig.~\ref{fig:Bi_dimer_dos}(e), Fig.~\ref{fig:Bi_dimer_dos}(h)). The same behavior has recently been observed for Pb dimers on a Fe bilayer on W(110)~\cite{Schoneberg2018}. 

%SH
In addition, the $p_x$ states are partially involved in the formation of molecular $\sigma$ bonds along the dimer axis and thereby not available for the mixing with the $p_z$ states which
further reduces the possible value of the TAMR. The reduction of the LDOS of $p_x$ states in the shown energy range due to hybridization
is even more apparent in the $[001]$ and $[1\bar{1}1]$ Bi dimers (cf.~Fig.~\ref{fig:Bi_dimer_dos}(c), Fig.~\ref{fig:Bi_dimer_dos}(i)) while it is similar to that of the Bi adatom (cf.~Fig.~\ref{fig:BiAdatom}) for the 
$[1\bar{1}0]$ dimer (cf.~Fig.~\ref{fig:Bi_dimer_dos}(f)).

For the $[001]$ Pb dimer, within the simplified model, one can interpret the changes in the curves of the minority $p_x$ and $p_z$ states upon rotation of the spin-quantization axis at $-0.52$ eV due to hybridization mediated by SOC (see Fig.~\ref{fig:Pb_dimer_dos}(c)). The same effect could already be observed for the single Pb adatom directly at {\eF} (cf. Fig.~\ref{fig:PbAdatom}(c)). 
If the dimer orientation ($[001]$) is perpendicular to the magnetization direction ($[1\overline{1}0]$) of the substrate, one would expect SOC to mix molecular $\pi_z$ and antibonding $\sigma^{*}$ orbitals which are composed of $p_x$ states here~\cite{Schoneberg2018}. Molecular orbitals of this symmetry are located further apart in the energy spectrum than $\pi_z$ and $\pi_x$ molecular orbitals that can easily hybridize via SOC for a dimer axis along the magnetization of the Mn/W(110) surface, {\ie} the $[1\overline{1}0]$ direction. Hence, within this simple model, the $[001]$ $6p$ dimers were expected to exhibit a much smaller variation of their electronic structure under the influence of SOC. However, our DFT calculations show that the anisotropy of the LDOS actually takes an unexpected high value of $-28$\% and $-37$\% for the case of the $[001]$ Pb and Bi dimer adsorbed on Mn/W(110), respectively. Hence, this model based on only two atomic/molecular states
%SH
is not sufficient to quantitatively understand the TAMR for dimers along the $[001]$ axis. 

\begin{figure}[htb]
	\centering
	\includegraphics[scale=0.85,clip]{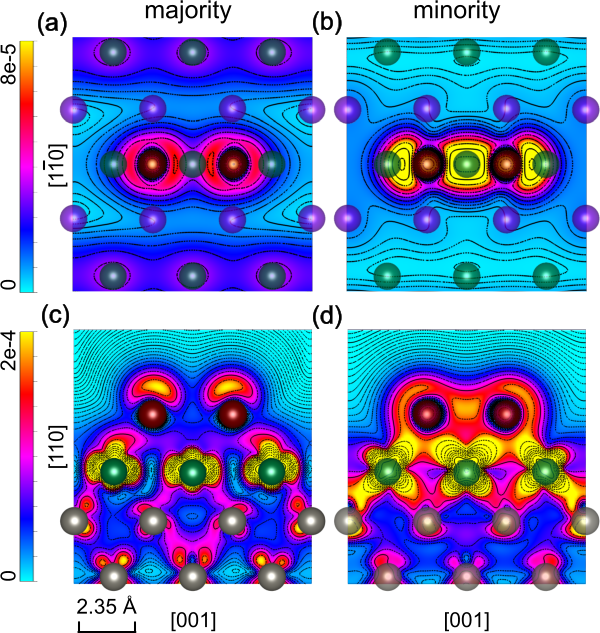}
	\caption{(a) and (b): top view (cross section) of the spin-resolved partial charge density plots of the Pb [001] dimer on Mn/W(110) in the energy range [{\eF}$-0.51$, {\eF}$-0.49$~eV]. (c) and (d): cross-sectional plots through the Pb dimer parallel to the [001] direction for the charge densities shown in (a) and (b).}
	\label{fig:pb_001_chg}
\end{figure}
\begin{figure}[htb]
	\centering
	\includegraphics[scale=0.85,clip]{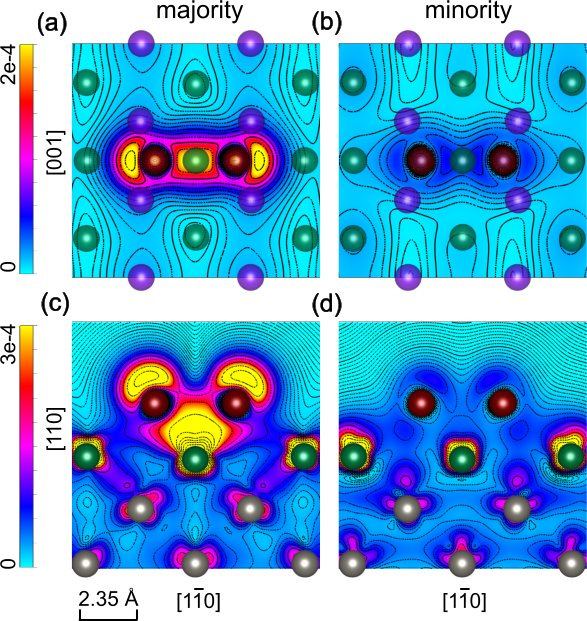}
	\caption{(a) and (b): top view (cross section) of the spin-resolved partial charge density plots of the Pb $[1\overline{1}0]$ dimer on Mn/W(110) in the energy range [{\eF}$-0.46$, {\eF}$-0.44$~eV]. (c) and (d): cross-sectional plots through the Pb dimer parallel to the $[1\overline{1}0]$ direction for the charge densities shown in (a) and (b). }
	\label{fig:pb_110_chg}
\end{figure}

In order to achieve a deeper understanding of the effect of $p-d$ hybridization on TAMR for the Pb and Bi dimers on Mn/W(110), we have calculated the partial charge densities  within the scalar-relativistic approximation,
{\ie} neglecting SOC, for a small energy range where the TAMR appears most prominent. The inclusion of SOC will not affect the  hybridization as evident from the LDOS in the scalar relativistic approximation [Fig.~\ref{fig:Pb_Bi_dimer_dos_full}] and including SOC [Figs.~\ref{fig:Pb_dimer_dos}, \ref{fig:Bi_dimer_dos}].
In the following we exemplify the influence of the substrate by means of both $[001]$ and $[1\overline{1}0]$ 
Pb dimers only as we observe similar characteristic behaviors for the Bi dimers. 

The spin-resolved partial charge density of the $[001]$ Pb dimer at approximately {\eF}$-0.5$ eV at which the large TAMR of $\sim$ 28\% occurs (cf. Fig.~\ref{fig:Pb_dimer_dos}(b)) is shown in Fig.~\ref{fig:pb_001_chg}. From the top view [Fig.~\ref{fig:pb_001_chg}(a,b)]
which represents a cross section through the dimer one can clearly see the molecular $\pi_z$ and $\sigma_x$ character of the adsorbate for both spin channels, whereas from the side view [Fig.~\ref{fig:pb_001_chg}(c,d)] a strong hybridization with the $d$ orbitals of the Mn atoms of the surface becomes visible.  
%SH
In the majority spin channel
the axes of the $p_z$ orbitals at the Pb dimer deviate from the $z$ direction of the unit cell and the $d$ orbitals of the Mn atoms are twisted towards the adsorbate [Fig.~\ref{fig:pb_001_chg}(c)]. 
This leads to a tilt of the upper lobes of the $p_z$ orbitals towards the center of the dimer and an overlap of the lower lobes with the Mn $d$ states in the case of the majority channel. 

For the minority channel on the other hand [Fig.~\ref{fig:pb_001_chg}(d)] a clear differentiation between the Pb and Mn states is not possible anymore due to the strong hybridization which becomes manifest in an accumulation of the charge density at the interface. These observations already indicate that for the explanation of the TAMR effect of the $6p$ dimers on Mn/W(110) more than two atomic states have to be taken into account.

Hybrid Pb-Mn interface states are also present in the majority channel of the $[1\overline{1}0]$ Pb dimer at the position of the maximum TAMR around 0.45 eV below {\eF} (see Fig.~\ref{fig:pb_110_chg}). A closer look at the calculated charge density reveals that its majority $p_z$ orbitals only interact with the Mn atom below the dimer axis, but not with the other atoms of the Mn monolayer [Fig.~\ref{fig:pb_110_chg}(c)]. Exactly the same behavior can also be observed for the corresponding $[1\overline{1}0]$ Bi dimer on Mn/W(110) (not shown). The reason for this hybridization can be explained by means of the different distances of the $6p$ atoms and their neighboring Mn atoms. While the central Mn atom and one atom of the Pb dimer are separated by just 2.84 {\AA}, the respective distance towards the next Mn atoms is 3.15 {\AA} and hence significantly larger. 

However, the interaction with the central Mn atom described above cannot be realized for the $[1\overline{1}1]$ dimer since the respective atom of the substrate is missing below a bonding axis along the diagonal (see Fig.~\ref{fig:geom}(c)).

\begin{figure}[htb!]
	\centering
	\includegraphics[scale=0.35,clip]{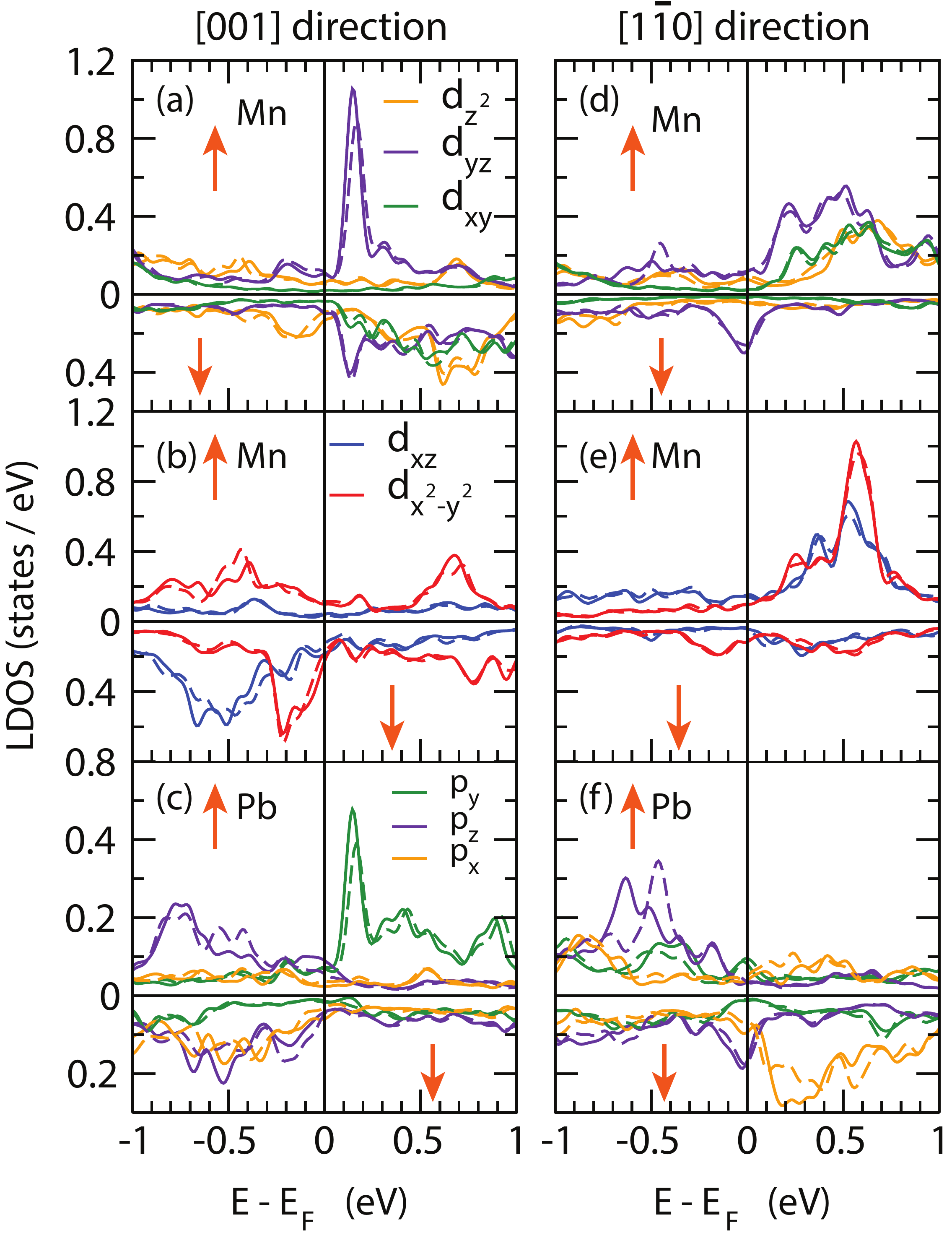}
	\caption{Orbital decomposed LDOS including SOC of the central Mn atoms below the $[001]$ and $[1\overline{1}0]$ Pb dimer in terms of the majority (up) and minority (down) states. Solid (dashed) lines correspond to the magnetization direction perpendicular (parallel) to the surface plane.  The orange up and down arrow indicates majority and minority spin channels, respectively.}
	\label{fig:Mn_Soc_Dos}
\end{figure}
Keeping in mind the studies of the partial charge densities, we did a further investigation of the LDOS of the central Mn atoms below the Pb dimer axes for the case of the $[001]$ and $[1\overline{1}0]$ direction. Fig.~\ref{fig:Mn_Soc_Dos} shows their orbitally decomposed $d$ states in an energy interval of $\pm$1 eV around {\eF} along with the $p$ states of the adsorbates for the two different magnetization directions discussed before.
It is evident that the Mn atoms below both Pb dimers are likewise affected by the rotation of the magnetization direction and bear resemblance to the changes in the states of the Pb atoms at same energetic positions. 

For the $[001]$ direction this becomes mostly apparent at 0.15 eV where a large peak of majority $d_{yz}$ orbitals of the central Mn atom shows the same behavior upon a change of the spin-quantization axis as the dominant $p_y$ state of the adsorbed dimer (cf. Fig.~\ref{fig:Pb_dimer_dos}(c)). Moreover one can observe an enhancement of the parallel magnetization component of majority states with $d_{z^2}$ and $d_{x^2-y^2}$ character at $-0.41$ eV which corresponds with a small peak of $n_{\parallel}$ for the majority $p_z$ states at the same energy. Further resemblance regarding magnetization-direction dependent differences in the LDOS are found between $-0.70$ eV and $-0.50$ eV for the minority $d_{xz}$ orbitals of Mn and $p_z$ and $p_x$ orbitals of the $[001]$ Pb dimer. 

For the $[1\overline{1}0]$ Pb dimer this SOC-dependent hybridization is most prominent for the $d_{yz}$ states both at {\eF} in the minority channel and at {\eF}$-0.45$ eV in the majority channel. Especially at the last-mentioned position, for the in-plane magnetization, it becomes clear that the majority Mn $d_{yz}$ states and the majority Pb $p_z$ states interacts quite 
strongly 
%SH
[Fig.~\ref{fig:Mn_Soc_Dos}] and produce a large TAMR value (see Fig.~\ref{fig:Pb_dimer_dos}(e)). 

\section{Conclusion}
In conclusion, we have presented a detailed study of the spin-resolved electronic structure of single Pb and Bi adatoms and dimers adsorbed on the Mn monolayer on W(110) including the
effect of spin-orbit coupling. 
Using density functional theory, we calculated the tunneling anisotropic magnetoresistance effect from two magnetization directions,
imposed due to the cycloidal spin spiral ground state in the Mn layer,  
for the respective $6p$ adsorbate: perpendicular to the surface (out-of-plane) and parallel to the $[1\overline{1}0]$ direction representing the propagation direction of the spin spiral ground 
state (in-plane). Our calculations for the $6p$ adatoms which are characterized by large spin-orbit coupling constants predict an enhancement of the TAMR up to 49\% for Pb
and 61\% for Bi adatoms. 

In both cases it can mainly be attributed to magnetization-direction dependent changes of majority $p_z$ states of the adatom. 
%SH not discussed in the paper: that strongly hybridize with the near neighbor Mn $d$ orbitals of the underlying Mn monolayer. 
The anisotropy of the LDOS of both adatoms can generally be explained by means of a simplified physical model which considers the coupling of two atomic states with different orbital symmetry 
($p_z$ and $p_x$ in the present case) via spin-orbit coupling. 
Although Pb and Bi adatoms carry almost no magnetic moment, they exhibit a large spin polarization directly at the surface and also in the vacuum due to the hybridization with the substrate. The spin polarization becomes maximal with values up to 60\% around {\eF} for the Pb adatom. 

We have also investigated the TAMR for three different dimer orientations adsorbed on the Mn/W(110) surface. 
Consistent with the expectations both Pb and Bi dimers with their bonding axis along the magnetization direction of the 
%SH
substrate, i.e.~the $[1\bar{1}0]$ direction,
show the maximum anisotropy of the vacuum LDOS with values of 64\% in the occupied regions. The origin of this large effect is a molecular $\pi_z$ orbital with majority spin character which strongly interacts with the central Mn atom below the dimer axis. Similar interactions 
are also found for a dimer orientation perpendicular to the magnetization direction of Mn/W(110), but with much smaller TAMR values of 37\% for Bi and 28\% for Pb, respectively. The TAMR becomes minimal for $6p$ dimers along the diagonal $[1\overline{1}1]$ direction (16\% in the case of Bi, 27\% for Pb) due to reduced SOC induced mixing of the $p$ states on the one hand and due to missing Mn atoms for hybridization below their bonding axes on the other hand. A further exploration of the central Mn atoms below the $[001]$ and $[1\overline{1}0]$ dimers has shown that their $d$ orbitals are likewise affected by changes upon rotation of the magnetization direction which has to be taken into account for the comprehension of the TAMR effect apart from the simple model of only two atomic states interacting by SOC. 

\section*{Acknowledgments} 
We acknowledge the DFG via SFB677 for financial support. We gratefully acknowledge the computing time at the supercomputer of the North-German Supercomputing Alliance (HLRN). We thank N. M. Caffrey for valuable discussions.

\end{document}